\documentclass[lettersize,journal]{IEEEtran}
\usepackage{amsmath,amsfonts}
\usepackage{algorithmic}
\usepackage{algorithm}
\usepackage{array}
\usepackage[caption=false,font=normalsize,labelfont=sf,textfont=sf]{subfig}
\usepackage{textcomp}
\usepackage{stfloats}
\usepackage{url}
\usepackage{verbatim}
\usepackage{graphicx}
\usepackage{cite}
\usepackage{booktabs}
\usepackage{multirow}
\usepackage{color}
\begin{document}

    \title{Two-Stage Radio Map Construction with\\Real Environments and Sparse Measurements}

\author{Yifan Wang,
Shu Sun,~\IEEEmembership{Member,~IEEE},
Na Liu,~\IEEEmembership{Member,~IEEE},
Lianming Xu,
and
Li Wang,~\IEEEmembership{Senior Member,~IEEE}
\thanks{%
This work was supported in part by the National Natural Science Foundation of China under Grants U2066201, 62101054, and 62271310, in part by the Fundamental Research Funds for the Central Universities under Grant No. 24820232023YQTD01, in part by the Double First-Class Interdisciplinary Team Project Funds 2023SYLTD06, and in part by Key Laboratory of Satellite Navigation Technology under Grant WXDHS2023105. (\emph{Corresponding author: L.~Wang}).}
\thanks{%
Y.~Wang, N.~Liu, L.~Xu, and L.~Wang are with the School of Computer Science (National Pilot Software Engineering School), Beijing University of Posts and Telecommunications, Beijing 100876, China (e-mail: wangyifan2019@bupt.edu.cn, na.liu@bupt.edu.cn, xulianming@bupt.edu.cn, liwang@bupt.edu.cn). }
\thanks{%
S.~Sun is with the Department of Electronic Engineering and the Cooperative Medianet Innovation Center (CMIC), Shanghai Jiao Tong University, Shanghai 200240, China (e-mail: shusun@sjtu.edu.cn)}}

\maketitle

\begin{abstract}
Radio map construction based on extensive measurements is accurate but expensive and time-consuming, while environment-aware radio map estimation reduces the costs at the expense of low accuracy. Considering accuracy and costs, a first-predict-then-correct (FPTC) method is proposed by leveraging generative adversarial networks (GANs). A primary radio map is first predicted by a radio map prediction GAN (RMP-GAN) taking environmental information as input. Then, the prediction result is corrected by a radio map correction GAN (RMC-GAN) with sparse measurements as guidelines. Specifically, the self-attention mechanism and residual-connection blocks are introduced to RMP-GAN and RMC-GAN to improve the accuracy, respectively. Experimental results validate that the proposed FPTC-GANs method achieves the best radio map construction performance, compared with the state-of-the-art methods. 
\end{abstract}


\section{Introduction}
\label{introduction}

Environment-aware radio map construction \cite{b1} has emerged as one of the most promising channel modeling techniques for next-generation wireless communication systems, which reflects the actual behaviors of wireless signals in specific propagation conditions \cite{b2}. A radio map typically refers to a graphical representation of a communication environment, capturing superposition effects of the environment-dependent wireless channel quality given fixed transmitters and communication frequency, e.g., the average received signal power \cite{b4}. Among the techniques for radio map construction, traditional methods such as empirical path loss models are computationally simple and fast but fail to predict accurately in complex propagation environments. Deterministic modeling methods such as ray tracing are proposed to improve the accuracy, but they are computationally expensive and time-consuming. 

To achieve satisfactory accuracy and speed, the latest works focus on utilizing artificial intelligence (AI). They can be categorized according to the acquired inputs: i) interpolation methods based on manual measurements, such as \cite{b10}. To ensure accuracy, a large number of manual measurements should be collected to gather the channel information from most areas, which requires substantial labor and time costs. ii) prediction methods based on environmental information, which can be obtained in advance or established by advanced real-time 3D environment sensing techniques. The frequently used environmental information includes the transmitter configuration \cite{b16}, obstacle locations \cite{b7}, obstacle heights \cite{b9}, and top/side view of the area \cite{b13}. However, the available environmental information is usually limited, which results in low accuracy. iii) methods using both environmental information and measurements, such as \cite{b18}. Although more valuable features can be extracted from both inputs, their collecting time is usually different. Environment information is generally obtained in advance with minor changes for a long time, while the measurements are collected after real-time activities, which leads to time inconsistency. The construction accuracy and efficiency will be compromised due to this problem.

To compensate for the deficiencies of existing methods, we propose a new structure called first-predict-then-correct (FPTC), to reduce the negative impact of time inconsistency. Moreover, two novel sub-networks are designed with different functional modules based on generative adversarial networks (GANs), thus achieving a radio map prediction GAN (RMP-GAN) and a radio map correction GAN (RMC-GAN). The RMP-GAN utilizes environmental information to predict coarse radio maps for basic applications, and the RMC-GAN further corrects the coarse results with sparse measurements to reflect real-time communication quality. Specifically, the RMP-GAN introduces a self-attention (SA) mechanism to capture long-distance environment dependencies, and the RMC-GAN adopts residual-connection (RC) blocks to preserve more valuable low-level physical information of the inputs.

\section{The Proposed FPTC-GANs Method}
\subsection{Overall Framework}
Considering an urban communication scenario $\Omega=\{\mathcal T, \mathcal O\}$, where multiple transmitters $\mathcal T = \{t_1, t_2, ..., t_T\}$ are deployed and obstacles $\mathcal O = \{o_1, o_2, ..., o_O\}$ with various densities are spread in the whole area. A small part of the whole area is considered as the interest area, defined as ${\Omega}_{ia} \in {\Omega}$, reserving partial transmitters ${\mathcal T}_{ia} \in {\mathcal T}$ and obstacles ${\mathcal O}_{ia} \in {\mathcal O}$ inside ${\Omega}_{ia}$. However, the reference signal received power (RSRP) inside the interest area may also be affected by the outside unknown environment $({\Omega}-{\Omega}_{ia})$, such as closely deployed transmitters and nearby obstacles, due to radio propagation characteristics. Hence, the true radio map $\mathbf{P}$ is expressed as 
\begin{equation}
\setlength{\abovedisplayskip}{3pt}
\setlength{\belowdisplayskip}{3pt}
\mathbf{P}= \xi({\Omega}_{ia}) + \mathbf{I}({\Omega}-{\Omega}_{ia}),
\end{equation}
where $\xi(\cdot)$ is the modelling function based on ${\Omega}_{ia}$, $\mathbf{I}({\Omega}-{\Omega}_{ia})$ is the influence induced by outside unknown environments. 

Complete environmental information of ${\Omega}_{ia}$ contains a lot of aspects such as the crowds and weather. To reduce the costs, ${\mathcal T}_{ia}$ and ${\mathcal O}_{ia}$ are considered as the accessible environmental information since they are relatively easy to obtain beforehand with minor changes for a long time. Deep-learning networks are adopted to better model $\xi(\cdot)$. A prediction network $f_{\theta}(\cdot)$ is used to predict coarse radio map $\mathbf{P}_{pre}$ by
\begin{equation}
\setlength{\abovedisplayskip}{3pt}
\setlength{\belowdisplayskip}{3pt}
\mathbf{P}_{pre}= f_{\theta}({\mathcal T}_{ia},{\mathcal O}_{ia}),
\end{equation}

Moreover, we have to remark that in most existing works, the $\mathbf{I}({\Omega}-{\Omega}_{ia})$ part is neglected ideally suggesting that the interest area is independent. However, there exists much influence in reality. Since $\mathbf{I(\cdot)}$ is unpredictable, sparse radio measurements $\mathcal M$ are considered to revise the coarse prediction closer to $\mathbf{P}$ via a correction network $g_{\theta}(\cdot)$. Additionally, the prediction error caused by other factors such as the unconsidered environment factors inside ${\Omega}_{ia}$ can also be corrected in this step. The corrected radio map $\mathbf{P}_{cor}$ is computed by
\begin{equation}
\setlength{\abovedisplayskip}{3pt}
\setlength{\belowdisplayskip}{3pt}
\mathbf{P}_{cor}= g_{\theta}(\mathbf{P}_{pre},\mathcal M).
\end{equation}

\begin{figure}[t!]
\setlength{\abovecaptionskip}{0pt}
\setlength{\belowcaptionskip}{10pt}
\centering
\graphicspath{{images/}}
\includegraphics[width=1\linewidth]{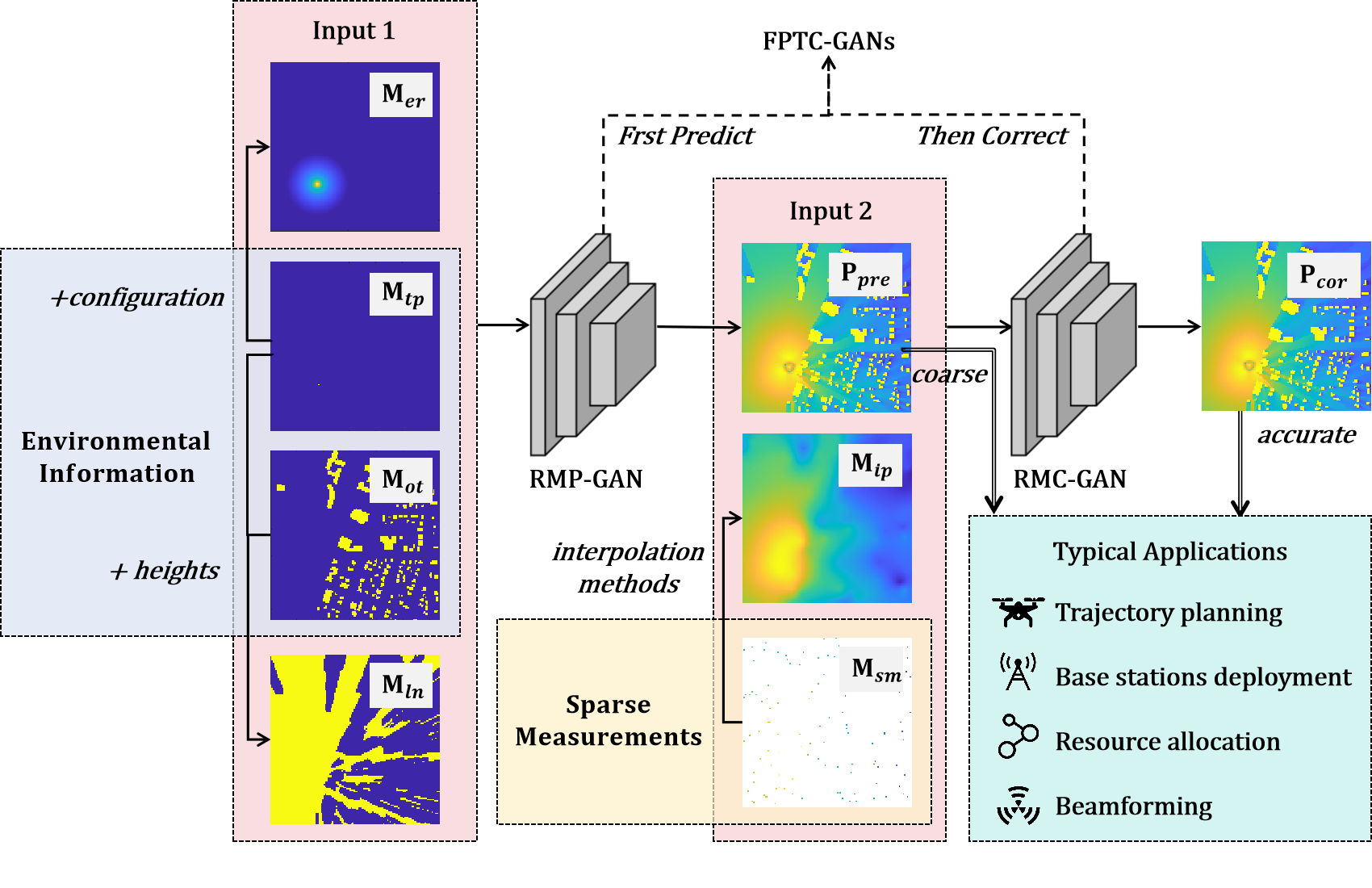}
\caption{Overall framework of the proposed FPTC-GANs method.}
\label{fig2}
\vspace{-1.5em}
\end{figure}

As clarified in Sec.~\ref{introduction}, the inconsistency of access time between the environmental information and measurements may lead to negative effects on radio map construction. A two-stage FPTC-GANs method is proposed to solve this problem. The overall framework is illustrated in Fig.~\ref{fig2}. This method follows a novel structure of FPTC, and it contains two sequentially connected sub-networks: an RMP-GAN $f_{\theta}(\cdot)$ and an RMC-GAN $g_{\theta}(\cdot)$. Specifically, the RMP-GAN is designed to process environmental information for coarse results $\mathbf{P}_{pre}$, followed by the RMC-GAN to correct $\mathbf{P}_{pre}$ with sparse measurements for more accurate results $\mathbf{P}_{cor}$. If the measurements are unavailable, $\mathbf{P}_{pre}$ can be directly used for basic applications. The proposed method has the advantage of obtaining the best possible results with all the available information, thus achieving a balance between efficiency, costs, and accuracy.

\subsection{Environmental Information and Sparse Measurements}
\label{dataconstruction}
To make the networks $f_{\theta}(\cdot)$ and $g_{\theta}(\cdot)$ better learn the mapping between inputs and outputs, the studied communication scenario ${\Omega}_{ia}$ is rasterized as 2D images to represent the inputs and outputs of the networks. The ${\mathcal T}_{ia}$ information is characterized by transmitter position $\mathbf{M}_{tp}$, height, transmit power, and frequency. The obstacle information ${\mathcal O}_{ia}$ is characterized by top view $\mathbf{M}_{ot}$ and obstacle heights. They are combined to generate empirical radio map $\mathbf{M}_{er}$ and LoS/NLoS indicator map $\mathbf{M}_{ln}$. The $\mathbf{M}_{tp}$, $\mathbf{M}_{ot}$, $\mathbf{M}_{er}$ and $\mathbf{M}_{ln}$ serve as the inputs for the RMP-GAN. The sparse measurements $\mathcal{M}$ are aggregated in $\mathbf{M}_{sm}$, and interpolated to become $\mathbf{M}_{ip}$. The $\mathbf{M}_{sm}$, $\mathbf{M}_{ip}$ and $\mathbf{P}_{pre}$ serve as the inputs for the RMC-GAN.

Empirical radio map $\mathbf{M}_{er}$ generated by the COST-231 Hata empirical model \cite{b14} is used to guide the training process of the RMP-GAN. It is suitable for this work due to the appropriate frequency band and application scenarios. Most of the existing works such as \cite{b9}\cite{b13}\cite{b12} perform calibration on the empirical models using the training set. Although it brings more accurate results, the generalization is reduced since much more data is required for calibration. To avoid this problem, we only use the basic parameters. The path loss PL is:
\setlength{\abovedisplayskip}{3pt}
\setlength{\belowdisplayskip}{3pt}
\begin{align}\label{eq}
\mathrm{PL}= &\ 46.3+33.9\log_{10}(f)-13.82\log_{10}(h_b)-ah_m\nonumber\\&+(44.9-6.55\log_{10}(h_b))\log_{10}(d)+c_m,
\end{align}
where $f$ is the frequency in MHz, and $h_b$ is the height of the transmitter above the ground in meters. For urban scenarios, $ah_m=3.20(\log_{10}(11.75h_r))^2-4.97$, where $h_r$ is the height of the receiver above the ground in meters. $c_m$ is 3 dB, and $d$ is the distance between the transmitter and receiver in kilometers.

LoS/NLoS indicator map $\mathbf{M}_{ln}$ is generated according to the relative positions between transmitters and receivers with consideration of the obstacles. If a direct LoS path exists, signal propagation follows the rule of free space path loss; otherwise, the signal will be blocked and propagation rules become complicated. Therefore, $\mathbf{M}_{ln}$ can help the RMP-GAN better learn the propagation modes. For each position in a radio map, it is an LoS state if $\forall h_l^i<h_{gmax}$, where $h_l^i$ represents the height of the $i$th point $\left(x_l^i,y_l^i\right)$ on the line $l$ between the transmitter $(x_b,y_b)$ and receiver $(x_r,y_r)$, $h_{gmax}$ represents the boundary value of LoS and NLoS paths calculated using geometry methods. Otherwise, it is an NLoS state. Assuming that the receivers are set lower than the transmitters, i.e., $\forall h_r<h_b$ as in most cases, $h_{gmax}$ is calculated as:
\begin{equation}
\setlength{\abovedisplayskip}{3pt}
\setlength{\belowdisplayskip}{3pt}
h_{gmax}=\ h_r+\frac{\sqrt{(x_l^i-x_r)^2+(y_l^i-y_r)^2}}{\sqrt{(x_b-x_r)^2+(y_b-y_r)^2}}\cdot|h_b-h_r|.
\label{eq}
\end{equation}

Sparse measurements are obtained from drive tests or crowdsensing \cite{b10}, which reflect the current situation on-site. Interpolated $\mathbf{M}_{ip}$ is generated quickly by performing simple interpolation methods on sparse measurements. The obtained $\mathbf{M}_{ip}$ is used as a guide for the RMC-GAN since the localized distribution trends around each measured point can be inferred. In this work, the radial basis function (RBF) \cite{b15} interpolation method is adopted since only a few points are known, leaving large blank areas. The calculated value $I_{x,y}$ for each data-free position $(x,y)$ is represented as:

\begin{equation}
I_{x,y}=\sum_{j=1}^N\phi\left(\sqrt{(x_j-x)^2+(y_j-y)^2}\right)w_j,
\label{eq}
\end{equation}
where $N$ is the total number of known samples, $(x_j,y_j)$ and $w_j$ are the position and weight of the $j$th known sample, respectively, and $\phi(\cdot)$ is the radial basis function.

\subsection{Networks Architecture}
\begin{figure*}[htbp]
\setlength{\abovecaptionskip}{0pt}
\setlength{\belowcaptionskip}{10pt}
\centering
\graphicspath{{images/}}
\includegraphics[width=0.9\linewidth]{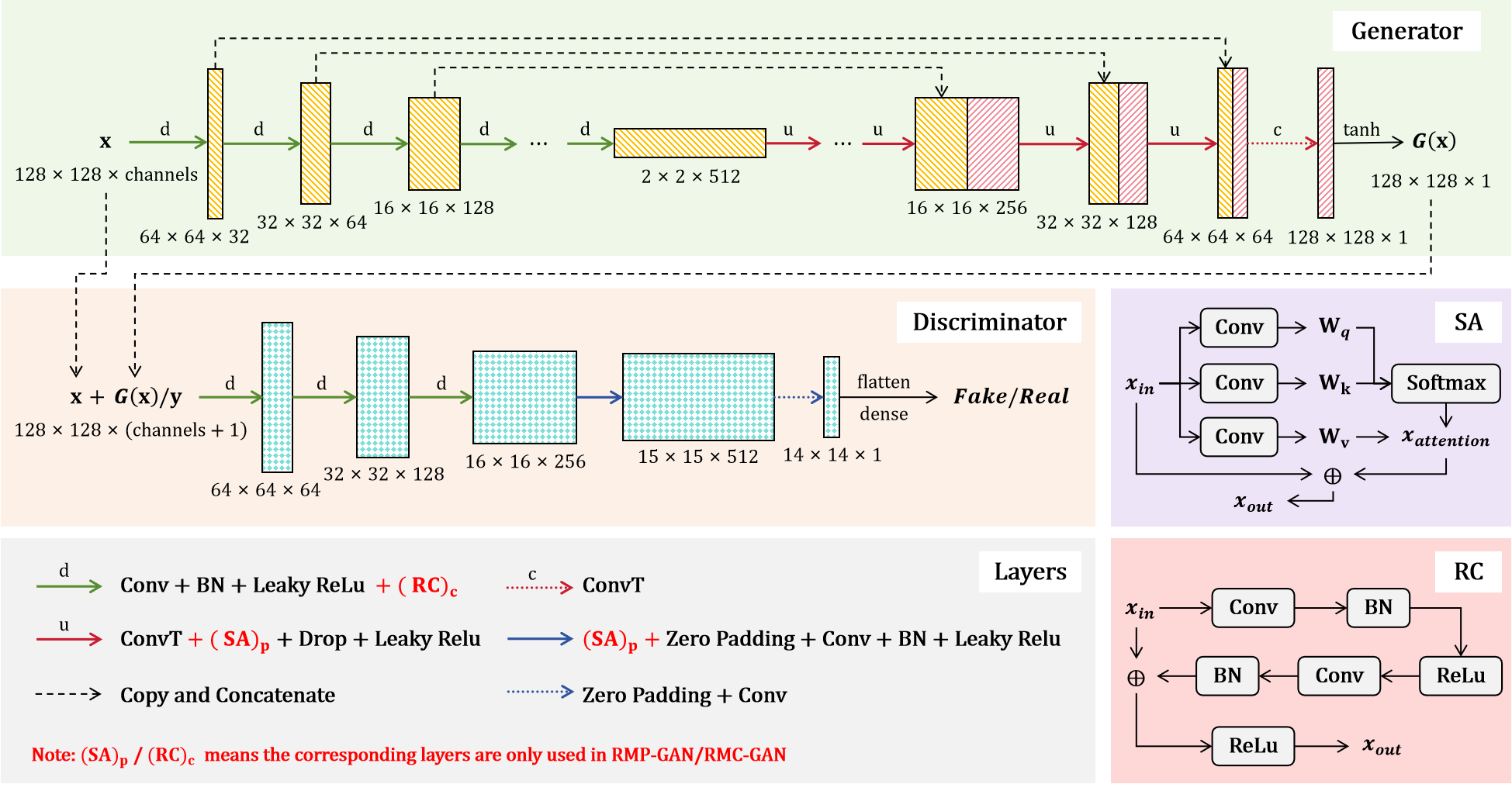}
\caption{The network architecture of the proposed RMP-GAN and RMC-GAN.}
\label{fig3}
\vspace{-1em}
\end{figure*}

As shown in Fig.~\ref{fig3}, the RMP-GAN and RMC-GAN are elaborated differently based on the conditional GAN (cGAN) \cite{b17}, which includes a generator $G$ and a discriminator $D$. Different functional modules, i.e., SA layers or RC blocks, are added respectively for the two networks.

The commonalities of the two networks are first introduced: For the generator, the inputs $\mathbf{x}$ are first downsampled through a convolution (Conv) layer, batch normalization (BN), and leaky rectified linear units activation (Leaky ReLu); then concentrated and upsampled through deconvolution (ConvT), dropout (Drop), and activated to generate $G(\mathbf{x})$. The discriminator distinguishes whether it is fake or real through downsampling structure, flattened and dense layers. As for the differences, SA layers are chosen for the RMP-GAN, and RC blocks are added for the RMC-GAN. The positions of these modules are highlighted using red color in Fig.~\ref{fig3}. The improvements caused by these modules are verified in Sec.~\ref{results}.

Since radio propagation is the mutual influence results of all obstacles in the entire scene, herein SA in the RMP-GAN helps understand spatial dependencies between long-distance locations, which contributes to capturing large-scale wireless channel features with a focus on key factors. Given the input feature $x_{in}$, SA enhances the feature extraction by
\begin{equation}
\setlength{\abovedisplayskip}{3pt}
\setlength{\belowdisplayskip}{3pt}
x_{out}=x_{in}+\gamma\cdot\left({softmax}\left(\mathbf{W}_q\cdot \mathbf{W}_k^{T}\right)\cdot \mathbf{W}_v\right),
\end{equation}
where $\mathbf{W}_q$, $\mathbf{W}_k$ and $\mathbf{W}_v$ are query, key, and value after $1 \times 1$ Conv layer; $\mathbf{W}_q$ and $\mathbf{W}_k$ are used to calculate attention scores through softmax. After that, it is multiplied by $\mathbf{W}_v$ for weighted result $x_{attention}$, which is scaled with a scalar $\gamma$ and added with the original $x_{in}$ to get the output $x_{out}$. 

RC blocks in the RMC-GAN preserve more valuable information about $\mathbf{P}_{pre}$ and sparse measurements, such that the model focuses on learning to correct $\mathbf{P}_{pre}$ rather than reconstructing the entire radio map from scratch. RC blocks operate through $x_{out}=Relu(x_{in}+x_{mid})$, where $x_{mid}$ is the intermediate features after Conv, BN, and ReLu. Then $x_{mid}$ is added with $x_{in}$ to conduct the final activation for $x_{out}$. 

The two sub-networks use the same ground truth to compute the loss with the output $\mathbf{P}_{pre}$ for RMP-GAN and $\mathbf{P}_{cor}$ for RMC-GAN, respectively. The loss function adopts adversarial loss and L2 distance loss. The adversarial loss is: 
\begin{equation}
\setlength{\abovedisplayskip}{3pt}
\setlength{\belowdisplayskip}{3pt}
\mathcal{L}_{ad}(G,D)=\mathbb{E}_{\mathbf{y}\sim p_{\mathrm{y}}}[\log D(\mathbf{y})]+\mathbb{E}_{\mathbf{x}\sim p_{\mathbf{x}}}[\log(1-D(G(\mathbf{x})))], 
\label{eq}
\end{equation}
where $p_{\mathbf{x}}$ and $p_{\mathbf{y}}$ stand for the distribution of input $\mathbf{x}$ and ground truth $\mathbf{y}$, respectively. The L2 distance loss is:
\begin{equation}
\setlength{\abovedisplayskip}{3pt}
\setlength{\belowdisplayskip}{3pt}
\mathcal{L}_{L2}=\mathbb{E}_{\mathbf{x},\mathbf{y}}\begin{bmatrix}\|\mathbf{x}-G(\mathbf{y})\|_2^2\end{bmatrix}. 
\label{eq}
\end{equation}

\section{Experiment Results and Analysis}
\subsection{Dataset}
The BART-Lab Radiomap Dataset\footnote{https://github.com/BRATLab-UCD/Radiomap-Data} is used to verify the proposed FPTC-GANs method. This dataset contains 2000 radio maps with different sizes varying from $140\times268$ pixels to $561\times1313$ pixels, each has corresponding top view maps and heights of the obstacles, as well as the positions and configuration of transmitters. Each map is cut to the interest areas with a fixed size of $128\times128$ pixels, preserving one transmitter and part of the obstacles inside the same map. In this work, 120 points (about 0.73\%) are randomly sampled from each radio map. Since we focus on outdoor scenarios, the area with obstacles is avoided. The inputs are generated as introduced in Sec.~\ref{dataconstruction}. 

After obtaining all inputs and corresponding target outputs, a dataset of 3518 groups of maps is created. The distribution density of obstacles and transmitter positions varies in each radio map. The dataset is separated into two training sets, two validation sets, and one testing set, to train the two sub-networks and compare their performance. The ratio of data volume in the five sub-datasets is $4:3:1:1:1$.

\subsection{Evaluation Criteria}
To evaluate the model performance, four quantitative measures are used for fair comparison in different aspects, which are root mean squared error (RMSE), mean absolute error (MAE), structural similarity (SSIM), and peak signal-to-noise ratio (PSNR). RMSE and MAE are used to evaluate the error between the construction result and target in dBm; SSIM and PSNR treat the radio maps as images with [0-255] pixel values to measure the similarity of the construction result and target. Moreover, the computational complexity is measured with the inference time in seconds and the number of parameters, where lower values stand for less computational complexity.

The RMSE between the ground truth $\mathbf{P}$ and the generated radio map $\widetilde{\mathbf{P}}$ is expressed as:
\begin{equation}
\setlength{\abovedisplayskip}{3pt}
\setlength{\belowdisplayskip}{3pt}
\mathrm{RMSE}(\mathbf{P},\widetilde{\mathbf{P}})=\sqrt{\frac1{N_p}\sum_{n=1}^{N_p}\left(\mathbf{P}_n-\widetilde{\mathbf{P}}_n\right)^2},
\label{eq}
\end{equation}
where $\mathbf{P}$ and $\widetilde{\mathbf{P}}$ both have $N_p$ signal-receiving points, the $n$th point are $\mathbf{P}_n$ and $\widetilde{\mathbf{P}}_n$, respectively. The MAE is expressed as:
\begin{equation}
\setlength{\abovedisplayskip}{3pt}
\setlength{\belowdisplayskip}{3pt}
\mathrm{MAE}(\mathbf{P},\widetilde{\mathbf{P}})=\frac1{N_p}\sum_{n=1}^{N_p}\left|\mathbf{P}_n-\widetilde{\mathbf{P}}_n\right|.
\label{eq}
\end{equation}
For the RMSE and MAE, lower values stand for better accuracy. The SSIM is expressed as:
\begin{equation}
\setlength{\abovedisplayskip}{3pt}
\setlength{\belowdisplayskip}{3pt}
\mathrm{SSIM}(\mathbf{P},\widetilde{\mathbf{P}})=\frac{(2\mu_P\mu_{\widetilde{P}}+C_1)(2\sigma_{P\widetilde{P}}+C_2)}{(\mu_P^2+\mu_{\widetilde{P}}^2+C_1)(\sigma_P^2+\sigma_{\widetilde{P}}^2+C_2)},
\label{eq}
\end{equation}
where $\mu_P$ and $\mu_{\widetilde{P}}$ are the local means of $\mathbf{P}$ and $\widetilde{\mathbf{P}}$, $\sigma_P^2$ and $\sigma_{\widetilde{P}}^2$ are the local variances of $\mathbf{P}$ and $\widetilde{\mathbf{P}}$, $\sigma_{P\widetilde{P}}$ is the local covariance between $\mathbf{P}$ and $\widetilde{\mathbf{P}}$. $C_1=(k_1L)^2$ and $C_2=(k_2L)^2$ are constants to stabilize the division when the denominator is close to zero, where $L=255$ for 8-bit images, ${k_1=0.01}$ and ${k_2=0.03}$. The PSNR is expressed as:
\begin{equation}
\setlength{\abovedisplayskip}{3pt}
\setlength{\belowdisplayskip}{3pt}
\mathrm{PSNR}\big(\mathbf{P},\widetilde{\mathbf{P}}\big)=10\log_{10}\frac{N_p(2^q-1)^2}{\sum_{n=1}^{N_p}\left(\mathbf{P}_n-\widetilde{\mathbf{P}}_n\right)^2},
\label{eq}
\end{equation}
where $q=8$ for 8-bit images. For the SSIM and PSNR, higher values stand for better similarity between the two images.

\subsection{Performance}
\label{results}
\begin{table}[!t]
\setlength{\tabcolsep}{1.8pt}
\centering
\caption{The performance of different models on the test set}
\label{tab1}
\begin{tabular}{cccccccc} 
\toprule 
 & Model & RMSE & MAE & SSIM & PSNR & Time & \#Parameters\\
\midrule 
\multirow{2}*{Prediction} &RMP-GAN &\textbf{3.18} &\textbf{2.36} &\textbf{0.86} &\textbf{26.03} & 0.03 & 20.0 M\\
            & cGAN & 3.40 & 2.57 & 0.83 & 25.37 &\textbf{0.01} &\textbf{19.5 M}\\
\midrule 
\multirow{2}*{Correction} &RMC-GAN &\textbf{2.03} &\textbf{1.35} &\textbf{0.90} &\textbf{29.45} & 0.02 & 25.8 M\\
            & cGAN & 2.25 & 1.56 & 0.86 & 28.52 &\textbf{0.01} &\textbf{19.5 M}\\
\midrule 
\multirow{5}*{Overall} &FPTC-GANs &\textbf{2.03} &\textbf{1.35} &\textbf{0.90} &\textbf{29.45} & 0.04 & 45.9 M\\
            & RadioUNet\cite{b7} & 2.55 & 1.83 & 0.89 & 26.88 & 0.02 & 33.5 M\\
            & GAN-CRME\cite{b18} & 2.79 & 2.02 & 0.85 & 26.03 & \textbf{0.01} & 19.5 M\\
            & Pix2Pix+\cite{b9} & 4.07 & 3.07 & 0.76 & 15.56 &\textbf{0.01} & 19.5 M\\
            & DeepREM\cite{b10} & 5.24 & 3.54 & 0.74 & 21.36 &\textbf{0.01} &\textbf{16.8 M}\\
\bottomrule 
\end{tabular}
\vspace{-1em}
\end{table}

Table~\ref{tab1} shows the average results of different methods. The effectiveness of extra SA layers is validated through the comparison between the RMP-GAN and cGAN in prediction tasks, and the results in correction tasks prove the capability of extra RC blocks (for fair comparison, both the RMC-GAN and cGAN in the correction part are realized using the same prediction results from the RMP-GAN). Both two novel networks show better accuracy than original cGANs, with acceptable increment in computational complexity.

Also, following the FPTC structure, the correction is carried out after the prediction. Comparing the results between the RMP-GAN and RMC-GAN, it is proved that the correction step is necessary since the accuracy is largely improved in all four aspects (RMSE, MAE, SSIM, and PSNR).

Moreover, the overall performance of the FPTC-GANs is compared with state-of-the-art works. RadioUNet \cite{b7} generates radio maps with the WNet. GAN-CRME \cite{b18} utilizes an original GAN with a cooperative radio map estimation (CRME) approach. Pix2Pix+ \cite{b9} is optimized based on the Pix2Pix, focusing on the difference between real radio maps and empirical models to constrain the range. To ensure fairness, the empirical model used in Pix2Pix+ is not calibrated, which can cause a slight decrease in accuracy. All these three methods utilize environmental information and measurements, but they do not follow the FPTC structure to separate the information. DeepREM \cite{b10} carries out interpolation using generative AI with 5\% measurements, which is much more than those used in other works. The results show that the proposed FPTC-GANs method outperforms other methods in accuracy. Although the computational complexity is slightly higher due to a second network and extra functional modules, the increment is acceptable compared with the advancement in accuracy, thus proving the effectiveness of this approach.

\begin{figure}[!b]
\vspace{-1em}
\setlength{\abovecaptionskip}{0pt}
\setlength{\belowcaptionskip}{10pt}
\centering
\graphicspath{{images/}}
\includegraphics[width=1\linewidth]{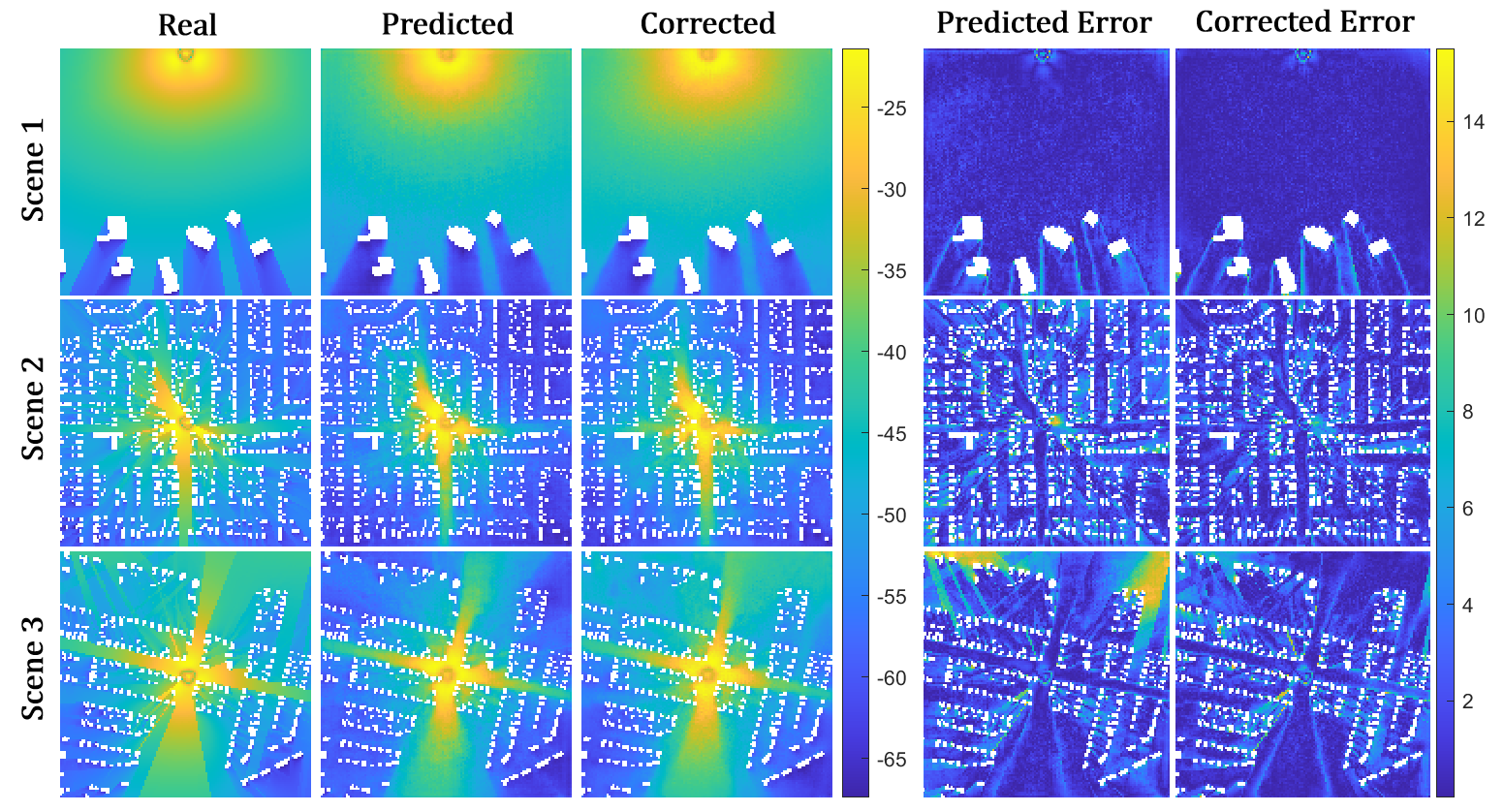}
\caption{Visualized radio maps (RSRP in dBm) for typical scenes.}
\label{fig5}
\end{figure}

Fig.~\ref{fig5} shows the radio maps in three typical scenes. By analyzing the results, it is found that for radio map prediction, the area with a higher density of obstacles usually results in lower accuracy. This can be confirmed in scenes 1 and 2, in which the former has better accuracy since the obstacle density is much lower. This may be due to simpler signal propagation rules and consistency with empirical models in open space, which is easier for the network to learn. As the density increases, more complicated propagation rules are involved, thus resulting in lower accuracy. Additionally, the higher influence level of the outside environment also results in more error. In scene 3, the upper left and right corners are significantly influenced by the outside environment, thus causing more error compared with other parts. This is probably because only the environmental information inside the target area is available, it is difficult for the model to learn from the unknown environment. As for radio map correction, the degree of improvement in accuracy after correction is proportional to the magnitude of prediction error. It is shown that scenes 1 and 2 have less prediction error than the last one, thus more error is corrected in scene 3. Therefore, radio map correction is especially important for scenes with low prediction accuracy.

\begin{figure}[!t]
\setlength{\abovecaptionskip}{0pt}
\setlength{\belowcaptionskip}{10pt}
\centering
\graphicspath{{images/}}
\includegraphics[width=0.9\linewidth]{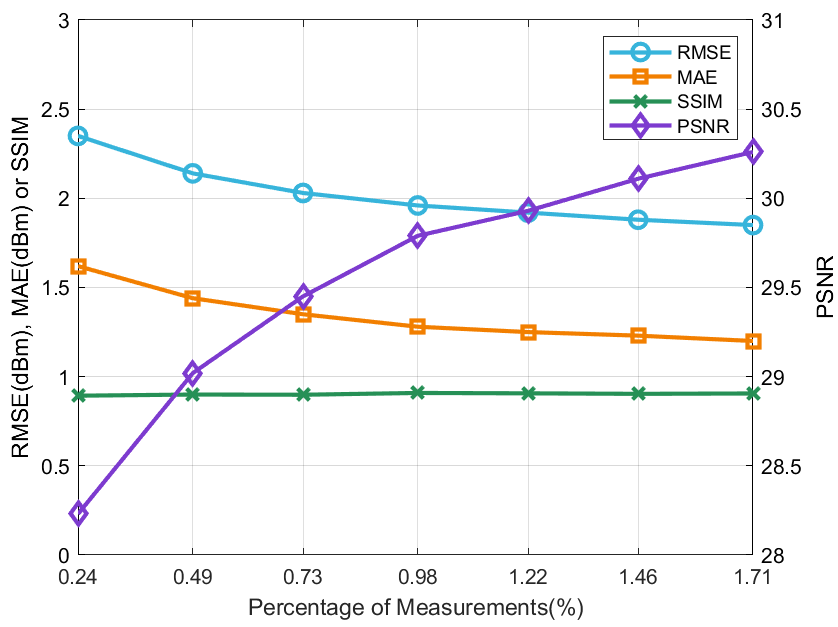}
\caption{RMC-GAN performance with different numbers of measurements.}
\label{fig7}
\vspace{-1em}
\end{figure}

Fig.~\ref{fig7} illustrates the performance of the RMC-GAN with different percentages of measurements. It is observed that more measurements result in better performance. However, it also requires more time/labor. The result gives the readers a guide to making a trade-off between accuracy and costs according to their needs in practical applications.

\begin{table}[!b]
\vspace{-1em}
\setlength{\tabcolsep}{6.7pt}
\centering
\caption{Models performance with different inputs on the test set}
\label{tab2}
\begin{tabular}{cccccc} 
\toprule 
 & Inputs & RMSE & MAE & SSIM & PSNR \\
\midrule 
\multirow{4}*{RMP-GAN} &\textbf{Original Inputs} &\textbf{3.18} &\textbf{2.36} &\textbf{0.86} & \textbf{26.03} \\
                      & Removing $\mathbf{M}_{er}$ & 3.52 & 2.70 & 0.80 & 25.01 \\
                      & Removing $\mathbf{M}_{ln}$ & 3.50 & 2.64 & 0.82 & 24.97 \\
                      & Removing $\mathbf{M}_{tp}$ & 3.34 & 2.52 & 0.82 & 25.55 \\
\midrule 
\multirow{2}*{RMC-GAN} &\textbf{Original Inputs} &\textbf{2.03} &\textbf{1.35} &\textbf{0.90} & \textbf{29.45} \\
                      & Removing $\mathbf{M}_{ip}$ & 2.14 & 1.46 & 0.90 & 29.00 \\
\bottomrule 
\end{tabular}
\end{table}

\subsection{Ablation Study}
Since there are multiple inputs for each network, it is necessary to verify the effect of each input. Following the idea of the ablation study, each input except the main inputs (information of obstacle top view and transmitter position in the RMP-GAN, coarse prediction and sparse measurements in the RMC-GAN) is removed separately to validate the importance of supporting information.

As shown in Table~\ref{tab2}, for the RMP-GAN, first the guidance provided by the empirical model and the LoS/NLoS information is verified, so $\mathbf{M}_{er}$ and $\mathbf{M}_{ln}$ are removed separately. Also, since $\mathbf{M}_{er}$ implicitly contains the approximate transmitter location, it is necessary to validate that $\mathbf{M}_{tp}$ is not redundant. As for the RMC-GAN, the guidance provided by the RBF interpolation needs to be confirmed, so $\mathbf{M}_{ip}$ is removed. The results show that the accuracy decreases for all these experiments. Therefore, it is proved that all the supporting information contributes to accurate radio map construction.

\section{Conclusion}
In this work, we presented a novel two-stage FPTC-GANs method to construct radio maps as close to reality as possible. The proposed method follows the FPTC structure and contains two generative deep-learning networks to carry out radio map prediction and correction tasks. Self-attention mechanism and residual-connection blocks are utilized respectively to adapt to the characteristics of corresponding tasks, thus yielding two novel models, the RMP-GAN and RMC-GAN, which can efficiently generate radio maps with high accuracy at low costs. The proposed FPTC-GANs method will benefit advanced emergency communication in the aspects of deterministic channel modeling, network deployment optimization, and autonomous vehicle trajectory planning.

\vfill

\end{document}